\pdfoutput=1
\documentclass[a4paper]{jpconf}
\usepackage{graphicx}
\begin{document}
\title{Cooling of suspended nanostructures with tunnel junctions}

\author{P.~J.~Koppinen and I.~J.~Maasilta}

\address{Nanoscience Center, Department of Physics, P.O.Box 35, FI--40014 University of Jyv\"askyl\"a, Finland}

\ead{panu.koppinen@jyu.fi, ilari.maasilta@jyu.fi}

\begin{abstract}
We have investigated electronic cooling of suspended nanowires with SINIS tunnel junction coolers. The suspended samples consist
of a free standing nanowire suspended by four narrow ($\sim$ 200 nm) bridges. We have compared two different cooler designs for cooling the suspended nanowire. We demonstrate that cooling of the nanowire is possible  with a proper SINIS cooler design.
\end{abstract}

\section{Introduction}
Reaching ultra-low temperatures is always a challenge, requiring complex techniques such as dilution refrigeration or adiabatic 
demagnetization. Especially demanding are nanoscale samples at sub-K temperatures, where tiny external heat loads ($\sim$ pW) can heat up the sample significantly due to the weakness of dissipation mechanisms \cite{jenniprl}. However, this weakness can be turned into a strength if one can think of a local cooling technique for the nanoscale sample. One clever way to perform local cooling is with the help of normal metal--insulator--superconductor (NIS) tunnel junctions \cite{nahum,pekola}.   

Most work on NIS coolers so far \cite{pekolareview} has concentrated on cooling small metallic islands on bulk substrates. In this case, the heat flow from the sample is limited by the electron-phonon interaction, resulting in mostly cooling of the electron gas and not the substrate itself. However, many applications, for example low-temperature detectors and mechanical resonators, would benefit significantly from cooling of both electrons and phonons. The phonons in thin membranes have been succesfully cooled with large area NIS junctions \cite{luukanen,clark2}, but cooling of phonon modes of a suspended nanowire has not been reported before, although it seems feasible based on theoretical estimates \cite{hekk}. 

\section{NIS coolers}
The basic principle of a NIS cooler  is based on the existence of the energy gap $\Delta$ of the superconductor. The forbidden 
states in the energy gap allow only electrons with energies higher than the gap to tunnel from normal metal into superconductor.
Hence, the energy gap acts as energy filter, which allows only hot electrons to be extracted from the normal metal island, 
resulting in cooling of the electron gas \cite{pekolareview}.
Heat transported by electrons through the junction can be calculated from the heat current equation
\begin{equation}\label{Pcool}
\dot{Q}=\frac{1}{e^2R_{T}}\int^{\infty}_{-\infty}(E-eV)g_{S}(E)[f_{N}(E-eV,T_{N})-f_{S}(E,T_{S})]dE
\end{equation} 
where $R_{T}$ is the tunneling resistance of the junction, V voltage across the junction, $g_{S}(E)$ density of states in the 
superconductor,$f_{N,S}$ Fermi--Dirac distributions and $T_{N,S}$ temperatures of normal metal and superconductor, respectively. 
The cooling power increases by a factor of two with two NIS junctions in series, i.e. in the case of a SINIS 
structure. Maximum cooling power of the SINIS cooler is close to the gap edge of the superconductor, i.e. best performance is 
obtained by voltage biasing the cooler at voltage $V \sim \Delta/e$. As such, this process cools only the electrons of the normal metal island directly. To be able to cool phonons in the vicinity of the junction, one needs to limit the phononic thermal conductance so much that it becomes a bottleneck for the heat flow. This could be achieved with thin SiN membranes, or even more strongly, with narrow beams as in this study.  

\begin{figure}[ht]
\begin{minipage}{22pc}
\includegraphics[width=22pc]{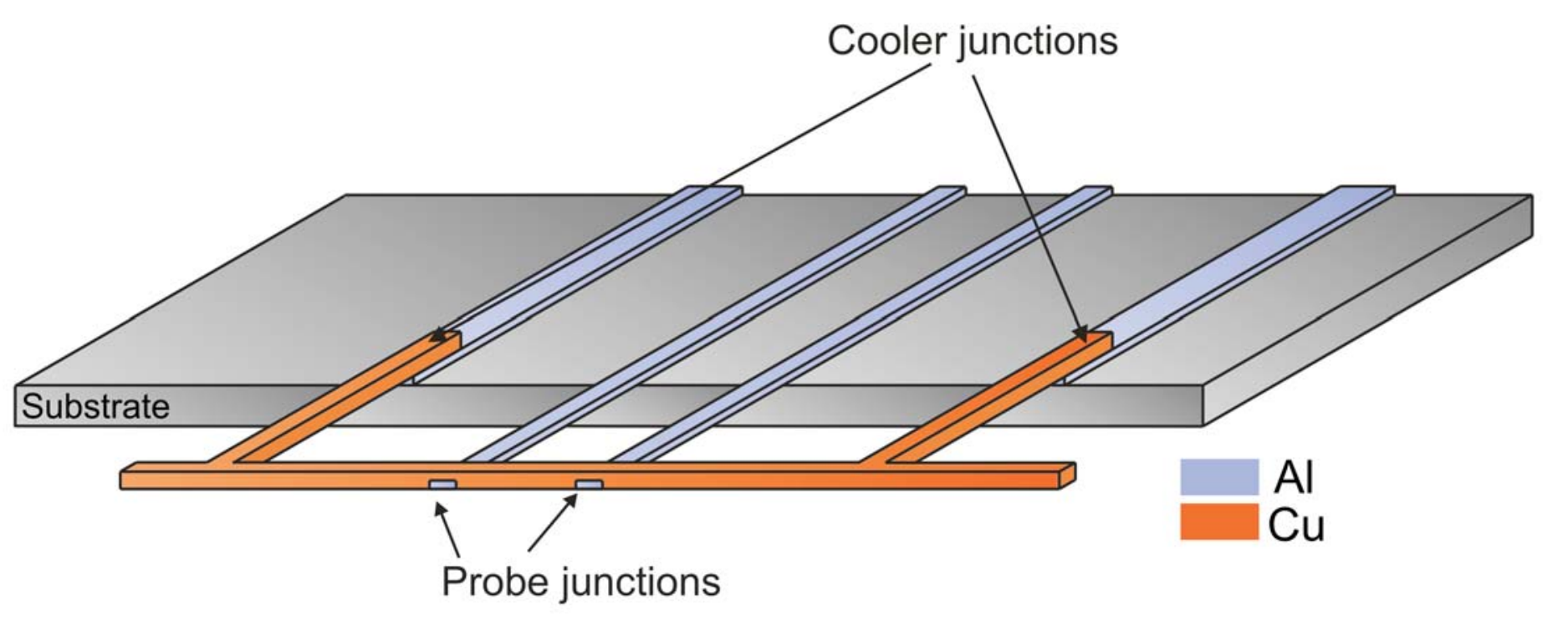}
\end{minipage}\hspace{2pc}%
\begin{minipage}{14pc}
\caption{\label{schema}(Color online) Schematic view of a suspended sample. Blue color represents aluminum and orange copper.}
\end{minipage}
\end{figure}

\section{Experimental}

The samples studied are suspended wires with lengths ranging between 10-20 $\mu$m, thickness 60 nm and width 150-300 nm. 
The nanowires have four supporting bridges (Fig. \ref{schema}) with typical dimensions of 150nm x 60 nm x 5 $\mu$m connecting it to the bulk, and limiting the phononic heat flow.

The sample design has two large area NIS tunnel junctions with junction area of 0.35 $\mu$m$^2$ as coolers, and two smaller (area 0.05 $\mu$m$^2$) NIS probe junctions used for thermometry \cite{row}. In addition, quasiparticle normal metal traps \cite{peko2} were fabricated near the cooler junctions to extract the tunnel current generated non-equilibrium quasiparticles from the superconducting leads. The quasiparticle traps were connected to the superconducting leads via a thin tunnel barrier. We also fabricated two different cooler geometries, with cooler junctions located either on the bulk (Fig. 1),  or suspended at the end of the supporting bridges. Probe junctions were always located at the free standing end of two middle bridges (Fig. \ref{schema}), measuring the temperature in the middle of the nanowire.  Typical tunneling resistances for coolers and probes were between 5-9 k$\Omega$ and 50-200 k$\Omega$, respectively.

All samples were fabricated on 30 nm thick low stress silicon nitride (SiN) LPCVD-grown on top of bulk silicon. SINIS junctions 
with Al (S) and Cu (N) electrodes with thickness of 30 nm were fabricated with conventional electron beam and shadow 
evaporation techniques in an ultra high vacuum (UHV) chamber. Before metal deposition, the substrate was cleaned with oxygen  
plasma to reduce the amount of organic impurities, e.g. resist residues on the substrate. Oxide barrier was grown in situ in the loading chamber of the evaporator by exposing the Al layer (evaporated first) to pure oxygen pressure of 10 mbar for 4 minutes. Oxidation was followed by deposition of 30 nm thick Cu film. The samples were fabricated on a 30 nm thick SiN membrane, which was released before the wire depositions by etching  a double-side nitridized (100) silicon wafer anisotropically in an aqueous KOH solution (bulk micromachining). A typical membrane size used was 50 $\mu$m $\times$ 50 $\mu$m. Finally, the SiN membrane was plasma etched with CHF$_3$ using the deposited metallic wires as etch masks, to release the final nanowire structure. This process thus leaves a suspended wire structure with 30 nm metallic top layer (Al or Cu) and a 30 nm insulating SiN bottom layer.   

After fabrication, the samples were measured in a $^3$He--$^4$He dilution refrigerator with a base temperature of $\sim$ 50 mK. 
The temperature of the nanowire was monitored with the probe junctions, by measuring the voltage of the junctions with constant 
current bias of $\sim$10 pA or $\sim$100 pA. This way, we obtain good responsivity for both low (T $<$ 500 mK) and high (500 mK 
$<$ T $<$ 1K) temperature range \cite{jenniphonons}. In the experiment, we simply sweep and measure the cooler junction voltage 
bias and see how temperature in the wire responds. 
 
\section{Results and discussion}
\begin{figure}[ht]
\begin{center}
\includegraphics[width=22pc]{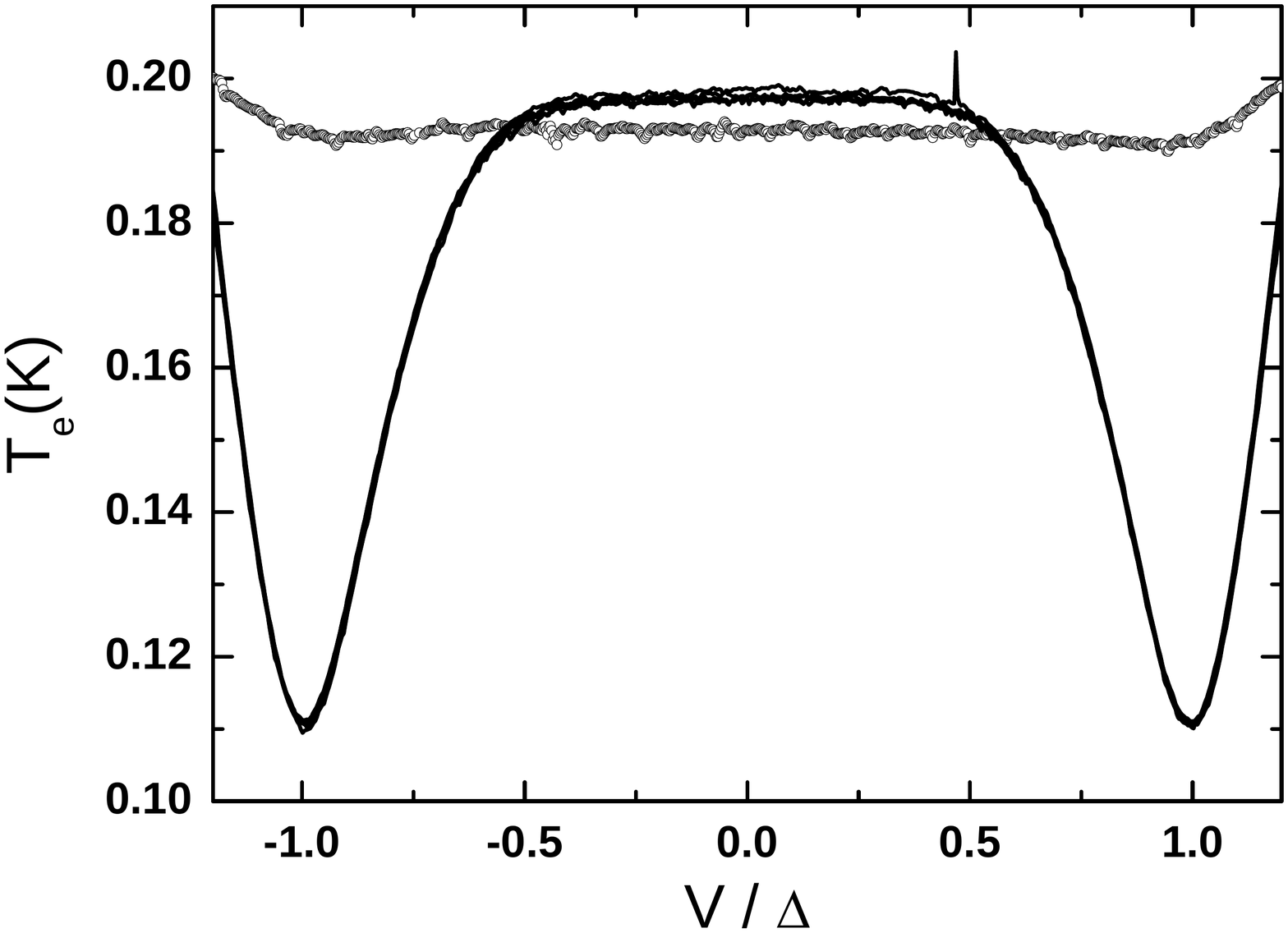}
\caption{\label{comparison}Electron gas $T_{e}$ temperature as function of cooler bias for two different suspended sample 
geometries. Open circles represent data for coolers located at the free standing end of the supporting bridges and solid line for geometry with cooler junctions on bulk.}
\end{center} 
\end{figure}

Figure \ref{comparison} shows the temperature of two suspended nanowire samples as a function of cooler bias voltage for the two different cooler geometries at a bath temperature of $\sim$200 mK. The open circles plot the data from the sample with cooler junctions suspended on the free standing end of the supporting bridges, whereas the solid line is the data for the sample with the cooler junctions on bulk substrate. The sample with cooler junctions on bulk shows cooling from 200 mK down to 110 mK, while  no significant cooling is observed in the sample with cooler junctions suspended. 

Differences in the cooling behavior can be explained with the reduced phonon thermal conductance of the supporting bridges. All 
hot non-equilibrium quasiparticles injected into the superconductor by the tunneling process eventually relax somewhere near the junction area via phonon emission. If these recombination phonons are generated in the suspended legs, they have a low rate of leakage to the bulk and thus high rate of reabsorption by the normal metal. This leads to no effective cooling, which is most likely the explanation for the weak cooling effect seen in the sample where the cooler junctions are also suspended (open circles, Fig. \ref{comparison}). However,  if the cooler junctions are located on the bulk, the recombination phonons are very likely emitted into the 3D substrate directly and thus not reabsorbed by the wire. This explains the better effectiveness of the second geometry (solid line, Fig. \ref{comparison}).   
 
\section{Conclusions}
In conclusion, electronic cooling of a suspended nanowire below the bath temperature is possible with normal-metal-superconductor  
tunnel junction coolers, if one uses a proper cooler design. Differences in the two designs discussed in this paper can be  
explained by considering the effect of the narrow suspended bridges on phonon transport. 
Undoubtedly, improvements in performance can be achieved by optimizing the sample parameters further. The observed cooling can  
open possibilities for application to cooling nanomechanical resonators to their ground states \cite{naik}, and future work will  
clarity further the cooling efficiency of our cooler design in a wider temperature range. 
Recently, tunnel junction cooling of a suspended nanowire was also demonstrated in a different cooler design \cite{muho}.

\ack
This work was supported by the Academy of Finland projects No. 118665 and 118231. P.J.K. acknowledges Magnus Ehrnrooth foundation  
for partial financial support.

\end{document}